%% file: ANA-HIGG-2021-21-PAPER.tex
\documentclass[cernpreprint, atlasdraft=false, USenglish, texmf, orcidlogo]{atlasdoc}
 
\usepackage{atlaspackage}
\usepackage{multirow}
 
\usepackage{atlasbiblatex}
 
\usepackage{atlasphysics}

\graphicspath{{logos/}{figures/}}
 
\usepackage{ANA-HIGG-2021-21-PAPER-defs}

\AtlasTitle{Determination of the Relative Sign of the Higgs Boson Couplings to $W$ and $Z$ Bosons Using $WH$ Production via Vector-Boson Fusion with the ATLAS Detector}

\AtlasAbstract{
The associated production of Higgs and $W$ bosons via vector-boson fusion is highly sensitive to the relative sign of the Higgs boson couplings
to $W$ and $Z$ bosons. In this Letter, two searches for this process are presented, using 140 fb$^{-1}$ of proton-proton collision data
at $\sqrt{s} = \SI{13}{\text{TeV}\xspace}$ recorded by the ATLAS detector at the LHC. The first search targets scenarios with opposite-sign couplings of the $W$ and $Z$ bosons to the Higgs boson,
while the second targets standard model-like scenarios with same-sign couplings. Both analyses consider Higgs boson decays into a pair of $b$ quarks and $W$ boson decays with
an electron or muon.
The data exclude the opposite-sign coupling hypothesis with a significance beyond $5\sigma$,
and the observed (expected) upper limit set on the
cross section for vector-boson fusion $WH$ production is 9.0 (8.7) times the standard model value at 95\% confidence level.
}

\AtlasRefCode{HIGG-2021-21}
 
\PreprintIdNumber{CERN-EP-2023-286}

\HepDataRecord{145856}
 
\AtlasJournalRef{\PRL 133 (2024) 141801}
\AtlasDOI{https://doi.org/10.1103/PhysRevLett.133.141801}

\hypersetup{pdftitle={ATLAS document},pdfauthor={The ATLAS Collaboration}}
 
\begin{document}
 
\maketitle

\clearpage
The study of the Higgs boson's couplings to $W$ and $Z$ bosons offers a crucial means of testing electroweak symmetry breaking
in the standard model (SM). These couplings can be parametrized in terms of the coupling modifiers $\kappa_{W}$ and $\kappa_{Z}$,
where values of 1 correspond to the SM expectation, or in terms of their ratio $\lam = \kappa_{W}/\kappa_{Z}$~\cite{Kappas}.
Any deviation of \lam{} from 1 would indicate a violation of the SM's custodial symmetry and be a clear sign of physics beyond
the standard model (BSM).
 
\begin{figure}[h]
\centering
\includegraphics[height=4.5 cm]{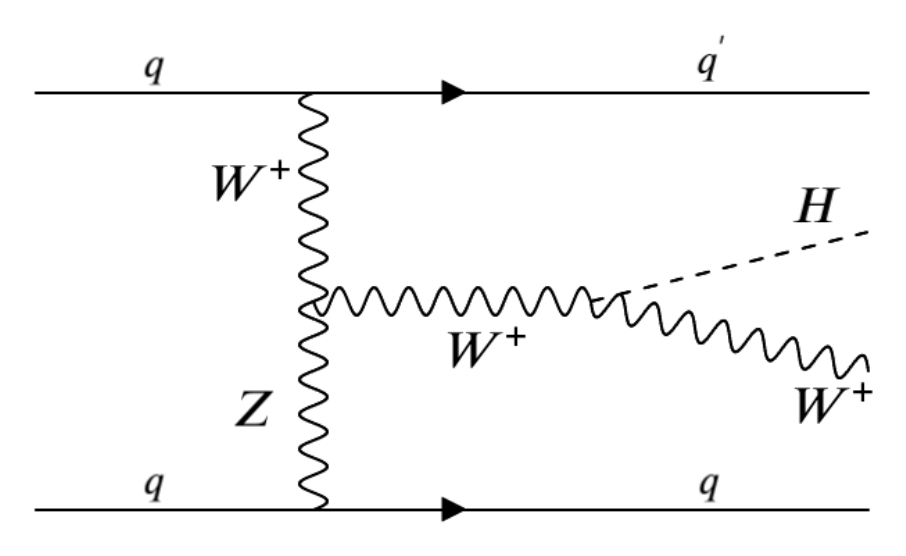}
\includegraphics[height=4.5 cm]{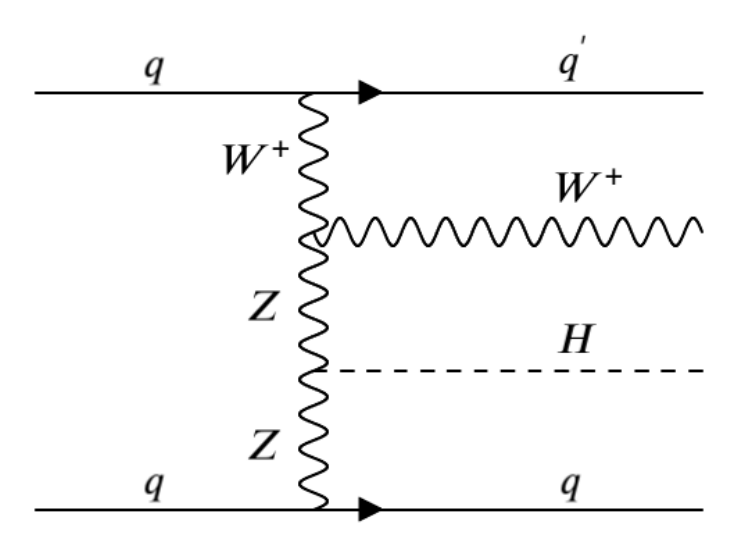}
\caption{
Examples of leading-order Feynman diagrams for VBF $WH$ production, where the Higgs boson interacts with either a $W$ or  $Z$~boson.
These diagrams interfere destructively if the Higgs boson couplings to $W$ and $Z$ have the same sign, and constructively if they have opposite sign.
\label{fig:Feyn}}
\end{figure}

By combining measurements of many Higgs boson production and decay modes, the ATLAS~\cite{HIGG-2021-23} and CMS~\cite{CMS-HIG-22-001}
collaborations have measured $|\lam|$ to be consistent with 1 with a precision of about 6\%. However, this relies primarily on
decays into $WW^{*}$ or $ZZ^{*}$, vector-boson fusion (VBF) production, and $WH$ and $ZH$ associated production, all of which scale with the squares of
$\kappa_{W}$ and $\kappa_{Z}$. Therefore, the relative sign of these parameters is nearly unconstrained by current measurements, and
they are both assumed to be positive in the coupling combinations.
Negative values of \lam{} are predicted by various models in which the observed Higgs boson is part of an isospin multiplet larger than a doublet~\cite{Low:2010jp},
as in the Georgi-Machacek model~\cite{GEORGI1985463}, making an experimental determination of its sign a key priority. In contrast
to those processes, the VBF $WH$ production mechanism~\cite{StoWu} includes diagrams where the Higgs boson couples to either a $W$ or $Z$ boson,
as shown in Fig.~\ref{fig:Feyn}. These diagrams interfere destructively in the SM, preserving unitarization of longitudinal gauge-boson scattering,
but the interference becomes constructive for negative values of \lam. This leads
to an enhancement in the cross section, particularly for events with large Higgs or $W$ boson momentum. Therefore, a measurement of this process can be used
to constrain the available ($\kappa_{Z}$, $\kappa_{W}$) parameter space with either the same or opposite sign.
Furthermore, the enhancement is due to tree-level interference, and therefore does not rely on assumptions regarding BSM loop contributions.
Other proposals to measure the sign of \lam{} include exploiting one-loop interference effects in $H \to 4\ell$ decays~\cite{Chen:2016ofc},
or using $W^{+}W^{-}H$ production~\cite{Chiang:2018fqf}.
 
This Letter presents two searches for VBF $WH$ production at the LHC, each using $140.1 \pm 1.2$ fb$^{-1}$~\cite{DAPR-2021-01} of
$\sqrt{s} = 13\,\TeV${} $pp$ collision data collected by the ATLAS detector during 2015--2018. The first search (\enquote{negative \lam}) targets BSM scenarios with a negative coupling ratio,
while the second (\enquote{positive \lam}) targets SM-like production.
Both analyses consider Higgs boson decays into $b\bar{b}$ and $W$ boson decays with an electron or muon (directly or via a $\tau$ lepton).
This leads to a final state with two $b$ jets, two additional jets from the protons, a charged lepton, and missing transverse
momentum (\met) from the neutrino(s).
 
The ATLAS experiment~\cite{PERF-2007-01} is a multipurpose particle detector with a forward-backward
symmetric cylindrical geometry and nearly \(4\pi\) coverage in solid angle~\cite{coordinates}.
It consists of an inner detector (ID) for tracking surrounded by a thin superconducting solenoid
providing a \qty{2}{\tesla} axial magnetic field, electromagnetic and hadron calorimeters, and a muon spectrometer.
A two-level trigger system is used to select events. The first-level trigger is implemented in hardware and uses a subset of the detector information
to accept events at a rate below \qty{100}{\kHz}. This is followed by a software-based trigger that reduces the accepted event rate to \qty{1}{\kHz} on average
depending on the data-taking conditions.
An extensive software suite~\cite{ATL-SOFT-PUB-2021-001} is used in data simulation, in the reconstruction
and analysis of real and simulated data, in detector operations, and in the trigger and data acquisition
systems of the experiment.
 
The VBF $WH$ process was simulated at leading-order accuracy in \alphas with \MGNLO~\cite{Alwall:2014hca} for the matrix element calculation, interfaced
to \PYTHIA[8]~\cite{Sjostrand:2014zea} for parton showering, hadronization, and multiple parton interactions.
The \NNPDF[3.0nlo] parton distribution function set~\cite{Ball:2014uwa} was used.
Predictions were obtained for various values of $\kappa_W$ and $\kappa_Z$ using the procedure outlined in Ref.~\cite{StoWu}.
Because the Lagrangian is insensitive to an overall sign change, only positive values of $\kappa_{W}$ were simulated.
The largest backgrounds come from \ttbar, $W$+jets, and $Wt$ single-top production, with smaller contributions from
$t$- and $s$-channel single-top, $Z$+jets, $VV$, $VH$, $t\bar{t}H$, and $t\bar{t}V$ production ($V$ = $W$ or $Z$).
Backgrounds from \ttbar{} and single top quark production were simulated with
\POWHEG~\cite{Frixione:2007nw, Nason:2004rx} interfaced to \PYTHIA[8]. Overlap between $Wt$ and \ttbar{} production was handled using the
diagram removal scheme~\cite{Frixione:2008yi}. The $W$+jets and $Z$+jets processes were simulated with \SHERPA[2.2.1]~\cite{Bothmann:2019yzt} for the matrix element and parton showering.
Different parton multiplicities were merged using the CKKW-L~\cite{Lonnblad:2001iq} technique.
Electroweak production of $WZ$ plus two jets was simulated at leading order with \MGNLO{} interfaced to \PYTHIA[8]. Other $VV$ processes were simulated
with \SHERPA, version 2.2.1 for quark-initiated processes and 2.2.2 for gluon-initiated processes. Other Higgs boson processes were generated with
\POWHEG, with the MiNLO procedure~\cite{Luisoni:2013cuh} applied for quark-induced $VH$, and with \PYTHIA[8] used for parton showering.
The $t\bar{t}V$ process was simulated at NLO with \MGNLO{} interfaced to \PYTHIA[8].
The background arising from misidentified or nonprompt leptons is evaluated using events with inverted lepton isolation requirements.
This background is seen to populate kinematic regions with low momentum objects and is found to be negligible after applying the final
selection criteria that define the signal and control regions.
 
Events in the electron channel were selected online using a single-electron trigger~\cite{TRIG-2018-05}. In the muon channel, events where the vector sum of the offline \met{} and the
muon \pt is greater than \SI{150}{\GeV} were selected with an \met{} trigger~\cite{TRIG-2019-01}, while below this threshold a single-muon trigger~\cite{TRIG-2018-01} was used.
(The trigger-level \met{} calculation does not include muons, making this vector sum a close approximation of the trigger-level \met.)
Because of the changing beam conditions, the kinematic and isolation requirements on the trigger objects varied during the data-taking period.
Electrons are reconstructed offline by matching clusters of energy deposits in the electromagnetic calorimeter to tracks in the ID, which are fitted allowing for energy loss
due to bremsstrahlung~\cite{EGAM-2018-01}. Events in the electron channel must have one electron candidate with $\pt{} > \SI{27}{\GeV}$ and $|\eta| < 2.47${}
passing the ``Tight'' likelihood identification criteria and the ``HighPtCaloOnly'' isolation criteria of Ref.~\cite{EGAM-2018-01}.
The electron must furthermore be matched to the primary vertex (the primary vertex is taken as the vertex with the highest sum of
squared transverse momenta of associated tracks) by requiring $|d_{0}|/\sigma_{d_{0}} < 5$ and $|z_{0}\sin(\theta)| < \SI{0.5}{\milli\meter}$, where $d_{0}${} is the
track's transverse impact parameter, $\sigma_{d_{0}}${} is its uncertainty, and $z_{0}$ is the longitudinal impact parameter. Muons are reconstructed
offline by matching tracks in the ID and muon spectrometer, accounting for energy loss in the calorimeters~\cite{MUON-2018-03}. Events in the muon channel must have
one muon satisfying $\pt > \SI{25}{\GeV}$ (\SI{27}{\GeV}) if an \met{} (a single-muon) trigger was used, $|\eta| < 2.5$, ``Medium'' quality, and
``HighPtTrackOnly'' isolation, as defined in Ref.~\cite{MUON-2018-03}. Similarly to electrons, the track must satisfy $|d_{0}|/\sigma_{d_{0}} < 3$
and $|z_{0}\sin(\theta)| < \SI{0.5}{\milli\meter}$. In both channels, events are rejected if a second lepton is present. For this veto, the \pt{} requirement
is lowered to \SI{7}{\GeV}, ``Loose'' identification and isolation requirements are applied~\cite{EGAM-2018-01,MUON-2018-03}, and the muon
pseudorapidity range is widened to $|\eta| < 2.7$.
 
Jets are reconstructed from particle-flow objects~\cite{PERF-2015-09}, combined using the anti-$k_t$~\cite{Cacciari:2008gp} algorithm with a radius
parameter of 0.4. Jets in the central region ($|\eta| < 2.5$) must have $\pt > \SI{20}{\GeV}$, while the \pt requirement is raised to \SI{30}{\GeV} for
forward jets ($2.5 < |\eta| < 4.5$). To reduce the effect of multiple collisions per bunch crossing, jets in the central (forward) region with $\pt < \SI{60}{\GeV}$ (\SI{120}{\GeV}) must have a
jet vertex tagger~\cite{ATLAS-CONF-2014-018} score $> 0.5$ (forward jet vertex tagger~\cite{ATL-PHYS-PUB-2019-026} score $< 0.5$). Jets in the central
region may be \enquote{$b$-tagged}, i.e.\ identified as containing a $b$-hadron decay, by combining information from sources such as
secondary-vertex reconstruction, track impact parameter measurements, and decay-chain fitting. The deep-learning algorithm DL1r~\cite{ATL-PHYS-PUB-2017-013,FTAG-2018-01}
is used, with a working point that has 70\% efficiency for $b$ jets from top quark decays.
The rates at which charm and light-flavor jets are incorrectly tagged as $b$ jets are around 10\% and 0.3\%, respectively~\cite{FTAG-2020-08,FTAG-2019-02}.
In addition to the standard jet calibration~\cite{PERF-2015-09},
two corrections are applied to $b$ jets to improve their energy resolution~\cite{HIGG-2016-29}. First, if any ``Medium''~\cite{MUON-2018-03} muons
with $\pt > \SI{5}{\GeV}$ and $|\eta| < 2.5$ are found within a cone of
jet-\pt-dependent size around the jet axis, the four-momentum of the closest muon is added to that of the jet. After this, a residual correction
is applied to equalize the response to jets with leptonic or hadronic decays of heavy-flavor hadrons. 
The \met{} is calculated as the negative vector sum of the transverse momenta of all jets and leptons in the event, plus a track-based soft term
accounting for other charged particles associated to the primary vertex~\cite{PERF-2016-07}.
 
Events must have exactly one charged lepton, exactly two $b$-tagged jets, and at least two nontagged jets. The two highest-\pt{} nontagged jets
are chosen as the VBF jets, and these are required to have a rapidity separation of $\dyjj > 3$.
After these requirements, approximately 430\;\!000 background events, primarily \ttbar, are
expected from simulation, compared to 860 signal events if $\kappa_W = 1$ and $\kappa_Z = -1$, or 50 if both parameter values are 1.
Selection criteria are applied to several kinematic variables to increase the signal-to-background ratio.
These include the VBF jets' invariant mass \mjj{} and rapidity separation \dyjj{},
as well as the $b$ jets' invariant mass \mbb{}, transverse momentum \pTbb, and angular separation \DRbb. 
The $W$ boson is reconstructed by summing the four-momenta of the lepton and neutrino, where the neutrino is assumed to have \pt{} equal to the observed \met{} and $\eta$
equal to that of the charged lepton. This is used to calculate the mass \mleptop{} of leptonically decaying top quarks, the centrality \centHW, and \DphiHWjj, according to the definitions
in Table~\ref{tab:vars}. Finally, \Nveto{} is defined as the number of jets with $\pt > \SI{25}{\GeV}$ and $|\eta| < 2.5$ which are not VBF or $b$-tagged
jets. In the negative-\lam{} analysis, a single signal region named \SRneg{} is used, while the positive-\lam{} analysis uses two orthogonal regions,
\SRposL{} and \SRposT, to enhance the sensitivity to the smaller SM signal. The selection criteria that define these regions are given in Table~\ref{tab:vars};
they were chosen to maximize the statistical significance, while keeping enough simulated events for robust estimation of the backgrounds and
systematic uncertainties. Compared to the negative-\lam{} signal, the SM signal has lower Higgs boson \pt, but similar VBF jet \pt{} and additional jet activity.
This motivates the higher \pTbb{} threshold in \SRneg, and the requirements on \mjj{} and \Nveto{} in \SRposL{} and \SRposT{}.
Distributions of the key kinematic observables used to define the signal regions are presented as Supplemental Material~\cite{supplemental}.
 
\begin{table}[htp]
\centering
\caption{
Definition of the signal regions used in the analyses. The SRs for the positive-\lam{} analysis are orthogonal: events in \SRposT{} are
excluded from \SRposL{}. The definition of the $W$ boson system is given in the text. 
}
\bgroup
\def\arraystretch{1.2}
\resizebox{\columnwidth}{!}{\begin{tabular}{ l l l l l l }
\toprule
Variable  & Description                               & \SRneg{}             & \SRposL{}            & \SRposT{} \\
\midrule
\mbb      & Invariant mass of the two $b$ jets ($b\bar{b}$ system).          & $\in(105,145)\,\GeV$ & $\in(105,145)\,\GeV$ & $\in(105,145)\,\GeV$ \\
\DRbb     & \dR{} between the two $b$ jets.              & $<1.2$               & $<1.6$               & $<1.2$ \\
\pTbb     & \pt{} of the $b\bar{b}$ system.              & $>\SI{250}{\GeV}$    & $>\SI{100}{\GeV}$    & $>\SI{180}{\GeV}$\\
\mjj      & Invariant mass of the VBF jets.            & --                   & $>\SI{600}{\GeV}$    & $>\SI{1000}{\GeV}$ \\
\Dyjj     & Rapidity separation of the VBF jets.       & $>4.4$               & $>3.0$               & $>3.0$             \\
\multirow{2}{*}{\mleptop}  & Invariant mass of the $W$ and either  & \multirow{2}{*}{$>\SI{260}{\GeV}$}    & \multirow{2}{*}{$>\SI{260}{\GeV}$}    & \multirow{2}{*}{$>\SI{260}{\GeV}$} \\
& $b$ jet that is closest to \SI{172.7}{\GeV} ($m_\mathrm{top}$).                     &                      &                      &       \\
\multirow{2}{*}{\centHW} & $\frac{|y_{Wb\bar{b}} - y_{jj}|}{\Dyjj}$, where $y_{Wb\bar{b}}$ ($y_{jj}$) is the & \multirow{2}{*}{$<0.3$}               & \multirow{2}{*}{$<0.3$}               & \multirow{2}{*}{$<0.3$} \\
& rapidity of the $Wb\bar{b}$ (VBF-jet) system. &                      &                      &       \\
\multirow{2}{*}{\DphiHWjj} & Azimuthal separation between the         & \multirow{2}{*}{--}                   & \multirow{2}{*}{--}                   & \multirow{2}{*}{$>2.7$}             \\
& $Wb\bar{b}$ system and the VBF-jet system.   &                      &                      &       \\
\multirow{2}{*}{\Nveto}    & Number of nontagged, non-VBF jets & \multirow{2}{*}{--}                   & \multirow{2}{*}{$\leq 1$}             & \multirow{2}{*}{$=0$}\\
&  with $\pt > \SI{25}{\GeV}$ and $|\eta| < 2.5$.   &                      &                      &       \\[1mm]
\bottomrule
\end{tabular}}
\egroup
\label{tab:vars}
\end{table}
 
In order to improve the background estimation, control regions (CRs) are defined for $\ttbar$, $W$+jets, and $Wt$,
separately for the two analyses. The CRs are dominated by the targeted background and depleted of signal, while maintaining key
kinematic features of the signal regions (SRs).
The definitions of the CRs are given in the Appendix.
For each analysis, the signal region(s) and the CRs are used together in a binned profile likelihood
fit~\cite{KendallStuart, Cowan:2010js}. The number of events in each region is taken as the observable. The normalization of each of the main
backgrounds is determined with an unconstrained parameter in the fit, \ktt, \kW, or \kWt, while systematic uncertainties are treated as
nuisance parameters with Gaussian or Poisson constraints.
 
Systematic uncertainties considered for the electrons and muons include those in the trigger, reconstruction, identification, and isolation efficiencies, and
the energy or momentum scale and resolution~\cite{EGAM-2018-01,MUON-2018-03}. For jets, uncertainties are considered for the energy
scale and resolution~\cite{JETM-2018-05}, the vertex tagging efficiency~\cite{ATLAS-CONF-2014-018,ATL-PHYS-PUB-2019-026}, and the
$b$-tagging efficiency for $b$ jets~\cite{FTAG-2018-01}, $c$ jets~\cite{FTAG-2020-08}, and light jets~\cite{FTAG-2019-02}. Uncertainties in the momenta
of all objects are propagated to the \met; additional uncertainties in the \met{} are considered for the soft term~\cite{PERF-2016-07} and for the trigger efficiency.
Systematic uncertainties in the modeling of the main backgrounds are assessed by replacing the nominal Monte Carlo (MC) predictions described previously with ones from
\MGNLO interfaced to \PYTHIA[8], and, for \ttbar and $Wt$, by using \HERWIG[7] as an alternative
parton shower algorithm. The treatment of overlap between \ttbar and $Wt$ is varied by using the alternative diagram subtraction scheme~\cite{White:2009}.
These uncertainties are symmetrized.
Additionally, the renormalization and factorization scales are varied by a factor of 2, and, for \ttbar and $Wt$, other parameters
sensitive to initial-state radiation are also varied~\cite{ATL-PHYS-PUB-2020-023}. Because the normalization of these backgrounds is
unconstrained in the likelihood fit, the systematic uncertainties affect only the relative contribution in each region. Uncertainties
in the cross section and acceptance of the smaller backgrounds are also considered, but their impact on the analysis is small. For the VBF $WH$
signal, the renormalization and factorization scales, the parton distribution function, and \alphas are varied~\cite{Butterworth:2015oua}.
In addition, an uncertainty in the modeling of additional jets in the positive-\lam{} analysis is assessed by using an alternative MC sample with
up to one additional parton in the matrix element, merged using the CKKW-L technique.
The total uncertainties of the signal yields in the SRs range from 11\% to 27\%.
 
Table~\ref{tab:numbers} presents the normalization factors and background yields in each SR obtained from the fit, as well as the predicted signal
and observed data yields. No significant pulls or constraints on any nuisance parameters are observed.
Because the CRs are not fully pure in the targeted background, the normalization factors for these backgrounds are anticorrelated by up to 71\% (for instance, a higher
value of \ktt{} would imply more \ttbar{} in the $W$+jets CR, and therefore a lower \kW; see the Supplemental Material~\cite{supplemental} for more information).
This results in the total predicted yield having an uncertainty smaller than the uncertainty in some individual components.
The postfit values of \ktt{} and \kW{} are close to unity, indicating good modeling by the simulation. The values of
\kWt{} are significantly below 1 because the MC prediction exceeds the data in the $Wt$ CRs. The phase space selected in
this analysis is highly sensitive to the treatment of \ttbar--$Wt$ overlap; when using the alternative diagram subtraction scheme, a deficit of MC events is seen in
the $Wt$ CRs, and values of \kWt{} close to 3 are obtained.
The difference between the baseline MC prediction and data is therefore smaller than the difference between the two MC predictions used to
estimate the systematic uncertainty. Moreover, because the $Wt$ normalization is determined from the fit to data,
the uncertainty in the measured signal strength from this source is less than 10\% of the total uncertainty.

\begin{table}[phtb]
\centering
\small
\sisetup{round-precision=2, group-digits=integer, group-minimum-digits=6}
\caption{
Normalization factors, expected background and signal yields, and observed data yields in each signal region. The background yields are given after the fit to data, while the
signal yields show both the prefit expectation and the fitted values. The VBF $WH$ signal corresponds to the prediction with $\kappa_{W} = 1$, $\kappa_{Z} = -1$
for \SRneg, and $\kappa_{W} = 1$, $\kappa_{Z} = 1$ for \SRposL{} and \SRposT{}.
The uncertainty in the total background is smaller than the quadrature sum of individual uncertainty components because of correlations.
}
\begin{tabular}{l |
S[table-alignment=right,table-format=4.1]@{\,$\pm$\,}S[table-alignment=left,table-format=2.1] |
S[table-alignment=right,table-format=4.2]@{\,$\pm$\,}S[table-alignment=left,table-format=2.2]
S[table-alignment=right,table-format=4.2]@{\,$\pm$\,}S[table-alignment=left,table-format=2.2]
}
\toprule
& \multicolumn{2}{c|}{Negative \lam} & \multicolumn{4}{c}{Positive \lam} \\
\midrule
\ktt                               &  \multicolumn{2}{c|}{$~0.88~~~^{+0.30}_{\text{--}0.35}$}    & \multicolumn{4}{c}{ $0.96~^{+0.21}_{\text{--}0.23}$ } \\ [1mm]
\kW                                &  \multicolumn{2}{c|}{$~1.12~~~^{+0.34}_{\text{--}0.25}$}    & \multicolumn{4}{c}{ $1.25~^{+0.33}_{\text{--}0.24}$ } \\ [1mm]
\kWt                               &  \multicolumn{2}{c|}{$~0.32~~~^{+0.39}_{\text{--}0.13}$}    & \multicolumn{4}{c}{ $0.31~^{+0.37}_{\text{--}0.14}$ } \\ [1mm]
$\mu = \sigma/\sigma_{\mathrm{pred.}}$ & \multicolumn{2}{c|}{$-0.027~^{+0.054}_{\text{--}0.057}$}  & \multicolumn{4}{c}{ $0.9\,~~^{+4.0}_{\text{--}4.3}~\,$ } \\ [1mm]
\midrule
& \multicolumn{2}{c|}{\SRneg{}}          &   \multicolumn{2}{c}{\SRposL{}}      & \multicolumn{2}{c}{\SRposT{}} \\
\midrule
\ttbar             & 42  &  19     & 172   &  35     & 15.0 &  5.8   \\
$W$+jets           & 26  &  13     & 84    &  32     & 14.1 &  7.6   \\
$Wt$               & 4.6 & 7.0     & 8     & 13      &  0.8 &  1.5   \\
Other background   & 5.4 &  1.6      & 16.2  & 4.2      & 3.0  & 1.5   \\
\midrule
Total background   & 77.7 & 8.6    & 279 & 15     & 32.9 & 5.8   \\
VBF $WH$, prefit  & 285  & 45      & 4.15 &  0.56  & 2.30 &  0.62 \\
VBF $WH$, postfit & \ensuremath{-8~~~}  & 17       & 4 &  17     & 2.2 &  9.8 \\
\midrule
Data               &  \multicolumn{2}{c|}{70}       &    \multicolumn{2}{c}{274}           &   \multicolumn{2}{c}{37}        \\
\bottomrule
\end{tabular}
\label{tab:numbers}
\end{table}
 
\begin{figure}[h]
\centering
\includegraphics[width=0.72\columnwidth]{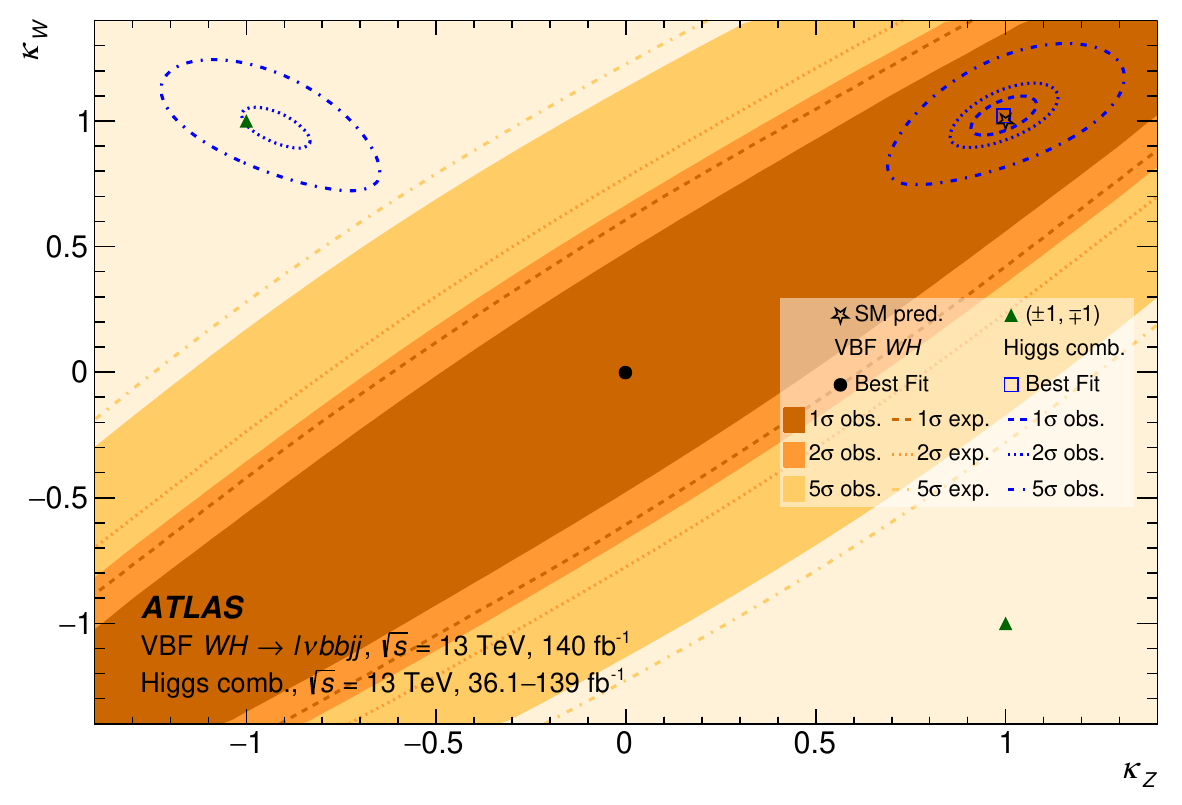}
\caption{
Fit results in the ($\kappa_{Z}$, $\kappa_{W}$) plane, using the negative-\lam{} analysis.
The results are overlaid with the confidence regions (shown in blue) from a separate fit using a combination of the Higgs boson measurements collected in Ref.~\cite{HIGG-2021-23}.
This fit assumes that all Higgs boson couplings besides the ones represented are positive, and that only SM particles contribute to loop processes.
Confidence regions are constructed from the log-likelihood ratio $\Lambda_{\text{LR}}=-2\ln(L/L_{\text{max}})$, where $L_{\text{max}}$ is the likelihood for the best-fit point,
which is shown as a black dot for the VBF $WH$ analysis or a blue dot for the Higgs combination.
The $1\sigma$, $2\sigma$, and $5\sigma$ regions are defined by $\Lambda_{\text{LR}}$ values smaller than 2.30, 6.18, and 28.7, respectively.
The SM value is marked with a star, while green triangles mark the points with $\kappa_{Z} = \pm1$, $\kappa_{W} = \mp1$.
\label{fig:2Dlimit}}
\end{figure}

In the negative-\lam{} analysis, 70 data events are observed in \SRneg{}, compared to an expectation of $80.6 \pm 8.6$ assuming the SM (including $2.93 \pm 0.35$ signal events), or $361 \pm 46$ in the
$\kappa_{W} = 1$, $\kappa_{Z} = -1${} scenario. (This is less than the sum of signal and background in Table~\ref{tab:numbers}, due to the effect that signal
contamination in the CRs would have on the background normalization factors.)
Figure~\ref{fig:2Dlimit} shows confidence regions in the $(\kappa_Z,\kappa_W)$ plane derived from the fit.
These are compared with the confidence regions obtained from a combination of
the ATLAS Higgs boson measurements collected in Ref.~\cite{HIGG-2021-23}.
This combination assumes that all coupling modifiers other than $\kappa_W$ and $\kappa_Z$ are positive. The resulting fit excludes
negative values of $\kappa_W$, primarily because of interference with the top quark in the loop decay $H \rightarrow \gamma\gamma$.
However, a region of parameter space with negative values of $\kappa_Z$ lies within one of the $2\sigma$ boundaries from the previous measurements.
This region is excluded by the present analysis with significance much greater than $5\sigma$. Thus the sign of \lam{} is determined to be positive.
 
In the positive-\lam{} analysis, 274 (37) events are observed in \SRposL{} (\SRposT{}), compared to an expected background
of $279 \pm 15$~($32.9 \pm 5.8$), and a SM signal of $4.15 \pm 0.56$~($2.30 \pm 0.62$). The fitted value of the signal strength $\mu = \sigma/\sigma_{\mathrm{pred.}}$ is
$0.9\,^{+2.3}_{-2.6}$~(stat.) $\,^{+3.3}_{-3.4}$~(syst.) $= 0.9\,^{+4.0}_{-4.3}$, indicating compatibility of the data with both the SM and  background-only hypotheses.
The largest systematic uncertainties come from the $W$+jets and \ttbar{} modeling, and the jet energy resolution.
An upper limit of 9.0 is set on the signal strength at 95\% confidence level (CL), compared to an expected limit of 8.7. The 95\% CL limit on the cross section for SM-like VBF $WH$ production
times the branching ratio $H \rightarrow b\bar{b}$ is 308~fb.
 
\FloatBarrier
 
In conclusion, the VBF $WH$ process has been studied by the ATLAS experiment, using \SI{140}{\per\femto\barn} of $pp$ collision data at $\sqrt{s} = \SI{13}{\TeV}$.
Events with two $b$ jets, a charged lepton, and two additional jets with a large rapidity gap are considered. No excess of events
is observed above the SM expectation. The 95\% CL upper limit set on the cross section for SM-like VBF $WH$ production is 9.0 times the SM prediction,
compared to an expected limit of 8.7. The $W$ and $Z$ boson couplings to the Higgs boson are determined to have the same sign, with previously unexcluded
opposite-sign hypotheses now excluded with significance beyond $5\sigma$.

\clearpage

\section*{Acknowledgments}


We thank CERN for the very successful operation of the LHC and its injectors, as well as the support staff at
CERN and at our institutions worldwide without whom ATLAS could not be operated efficiently.
 
The crucial computing support from all WLCG partners is acknowledged gratefully, in particular from CERN, the ATLAS Tier-1 facilities at TRIUMF/SFU (Canada), NDGF (Denmark, Norway, Sweden), CC-IN2P3 (France), KIT/GridKA (Germany), INFN-CNAF (Italy), NL-T1 (Netherlands), PIC (Spain), RAL (UK) and BNL (USA), the Tier-2 facilities worldwide and large non-WLCG resource providers. Major contributors of computing resources are listed in Ref.~\cite{ATL-SOFT-PUB-2023-001}.
 
We gratefully acknowledge the support of ANPCyT, Argentina; YerPhI, Armenia; ARC, Australia; BMWFW and FWF, Austria; ANAS, Azerbaijan; CNPq and FAPESP, Brazil; NSERC, NRC and CFI, Canada; CERN; ANID, Chile; CAS, MOST and NSFC, China; Minciencias, Colombia; MEYS CR, Czech Republic; DNRF and DNSRC, Denmark; IN2P3-CNRS and CEA-DRF/IRFU, France; SRNSFG, Georgia; BMBF, HGF and MPG, Germany; GSRI, Greece; RGC and Hong Kong SAR, China; ISF and Benoziyo Center, Israel; INFN, Italy; MEXT and JSPS, Japan; CNRST, Morocco; NWO, Netherlands; RCN, Norway; MEiN, Poland; FCT, Portugal; MNE/IFA, Romania; MESTD, Serbia; MSSR, Slovakia; ARRS and MIZ\v{S}, Slovenia; DSI/NRF, South Africa; MICINN, Spain; SRC and Wallenberg Foundation, Sweden; SERI, SNSF and Cantons of Bern and Geneva, Switzerland; MOST, Taipei; TENMAK, T\"urkiye; STFC, United Kingdom; DOE and NSF, United States of America.
 
Individual groups and members have received support from BCKDF, CANARIE, CRC and DRAC, Canada; CERN-CZ, PRIMUS 21/SCI/017 and UNCE SCI/013, Czech Republic; COST, ERC, ERDF, Horizon 2020, ICSC-NextGenerationEU and Marie Sk{\l}odowska-Curie Actions, European Union; Investissements d'Avenir Labex, Investissements d'Avenir Idex and ANR, France; DFG and AvH Foundation, Germany; Herakleitos, Thales and Aristeia programmes co-financed by EU-ESF and the Greek NSRF, Greece; BSF-NSF and MINERVA, Israel; Norwegian Financial Mechanism 2014-2021, Norway; NCN and NAWA, Poland; La Caixa Banking Foundation, CERCA Programme Generalitat de Catalunya and PROMETEO and GenT Programmes Generalitat Valenciana, Spain; G\"{o}ran Gustafssons Stiftelse, Sweden; The Royal Society and Leverhulme Trust, United Kingdom.
 
In addition, individual members wish to acknowledge support from Chile: Agencia Nacional de Investigaci\'on y Desarrollo (FONDECYT 1190886, FONDECYT 1210400, FONDECYT 1230812, FONDECYT 1230987); China: National Natural Science Foundation of China (NSFC - 12175119, NSFC 12275265, NSFC-12075060); Czech Republic: PRIMUS Research Programme (PRIMUS/21/SCI/017); EU: H2020 European Research Council (ERC - 101002463); European Union: European Research Council (ERC - 948254), Horizon 2020 Framework Programme (MUCCA - CHIST-ERA-19-XAI-00), European Union, Future Artificial Intelligence Research (FAIR-NextGenerationEU PE00000013), Italian Center for High Performance Computing, Big Data and Quantum Computing (ICSC, NextGenerationEU), Marie Sklodowska-Curie Actions (EU H2020 MSC IF GRANT NO 101033496); France: Agence Nationale de la Recherche (ANR-20-CE31-0013, ANR-21-CE31-0013, ANR-21-CE31-0022), Investissements d'Avenir Idex (ANR-11-LABX-0012), Investissements d'Avenir Labex (ANR-11-LABX-0012); Germany: Baden-Württemberg Stiftung (BW Stiftung-Postdoc Eliteprogramme), Deutsche Forschungsgemeinschaft (DFG - 469666862, DFG - CR 312/5-1); Italy: Istituto Nazionale di Fisica Nucleare (FELLINI G.A. n. 754496, ICSC, NextGenerationEU); Japan: Japan Society for the Promotion of Science (JSPS KAKENHI JP21H05085, JSPS KAKENHI JP22H01227, JSPS KAKENHI JP22H04944, JSPS KAKENHI JP22KK0227); Netherlands: Netherlands Organisation for Scientific Research (NWO Veni 2020 - VI.Veni.202.179); Norway: Research Council of Norway (RCN-314472); Poland: Polish National Agency for Academic Exchange (PPN/PPO/2020/1/00002/U/00001), Polish National Science Centre (NCN 2021/42/E/ST2/00350, NCN OPUS nr 2022/47/B/ST2/03059, NCN UMO-2019/34/E/ST2/00393, UMO-2020/37/B/ST2/01043, UMO-2021/40/C/ST2/00187); Slovenia: Slovenian Research Agency (ARIS grant J1-3010); Spain: BBVA Foundation (LEO22-1-603), Generalitat Valenciana (Artemisa, FEDER, IDIFEDER/2018/048), La Caixa Banking Foundation (LCF/BQ/PI20/11760025), Ministry of Science and Innovation (MCIN \& NextGenEU PCI2022-135018-2, MICIN \& FEDER PID2021-125273NB, RYC2019-028510-I, RYC2020-030254-I, RYC2021-031273-I, RYC2022-038164-I), PROMETEO and GenT Programmes Generalitat Valenciana (CIDEGENT/2019/023, CIDEGENT/2019/027); Sweden: Swedish Research Council (VR 2018-00482, VR 2022-03845, VR 2022-04683, VR grant 2021-03651), Knut and Alice Wallenberg Foundation (KAW 2017.0100, KAW 2018.0157, KAW 2018.0458, KAW 2019.0447); Switzerland: Swiss National Science Foundation (SNSF - PCEFP2\_194658); United Kingdom: Leverhulme Trust (Leverhulme Trust RPG-2020-004); United States of America: U.S. Department of Energy (ECA DE-AC02-76SF00515), Neubauer Family Foundation.


\clearpage
\appendix
\part*{Appendix}
\addcontentsline{toc}{part}{Appendix}

In order to improve the background estimation, control regions (CRs) are defined for $\ttbar$, $W$+jets, and $Wt$,
separately for the two analyses. The CRs are designed to be dominated by the targeted background and depleted of signal, while maintaining important
kinematic features of the signal regions.
Table~\ref{tab:CRs} presents the definitions of the CRs for each of the two analyses.
The \ttbar{} CRs use the high \mbb{} sideband, and consider values of \dyjj{} or \mjj{} that are lower
than in the SRs. The \ttbar{} events with high \mleptop{} often have a misidentified charm jet in place of the $b$-jet from leptonic top quark decay;
to preserve this feature, the \ttbar{} CRs maintain a minimum threshold for this variable. The $W$+jets CRs use a two-dimensional cut on
\drbb{} and \pTbb{} to find events where the $b$-jets are too close together to be consistent with the Higgs boson mass. The $Wt$ CRs
use high values of \drbb{} to remove signal, and high values of \mleptop{} and $W$ boson
\pt to reduce the contamination from \ttbar{} events.

\begin{table} [!htp]
\centering
\caption{
Definitions of the control regions for \ttbar, $W$+jets, and $Wt$. The $W$ boson's transverse momentum \ptw{} is the vector sum of the lepton \pT{} and
\met; the $W$ boson's transverse mass is calculated as $\mTW = \sqrt{2\,\met\,\pT^\ell\,(1-\cos{\phi})}$, where $\phi$ is the azimuthal angle between
the lepton and \met; \ptlj{} is the \pT{} of the leading VBF jet. Other variables are defined in Table~\ref{tab:vars}.
}
\bgroup
\resizebox{\columnwidth}{!}{\begin{tabular}{ l l l l l l l l }
\toprule
Variable   &  \ttbar{} CR$^-$      & \ttbar{} CR$^+$        &  $W$+jets CR$^-$       & $W$+jets CR$^+$       &  $Wt$ CR$^-$        &  $Wt$ CR$^+$         \\
\midrule
\mbb       &  $>\SI{145}{\GeV}$    & $>\SI{145}{\GeV}$      &  $<\SI{70}{\GeV}$      & $<\SI{70}{\GeV}$      & $>\SI{145}{\GeV}$   &  $>\SI{145}{\GeV}$   \\
\drbb      &  $<1.2$               & $<1.2$                 &  $< 2.23 - 0.007\pTbb/{\GeV}$ & $< 2.23 - 0.007\pTbb/{\GeV}$ & $>1.5$              &  $>1.6$              \\
\pTbb      &  $>\SI{200}{\GeV}$    & --                     &  $\in (150,250)\,\GeV$ & $>\SI{80}{\GeV}$      & $>\SI{250}{\GeV}$   &  $>\SI{180}{\GeV}$   \\
\mleptop   &  $>\SI{260}{\GeV}$    & $>\SI{220}{\GeV}$      &  $>\SI{275}{\GeV}$     & $>\SI{260}{\GeV}$     & $>\SI{320}{\GeV}$   &  $>\SI{320}{\GeV}$   \\
\Dyjj      &  $\in (3,4.4)$        & $>3$                   &  $>3$                  & $>3$                  & $>3$                &  $>3$                \\
\mjj       &  --                   & $\in (400,1000)\,\GeV$ &  --                    & $>\SI{500}{\GeV}$     & --                  &  $>\SI{500}{\GeV}$   \\
\Nveto     &  --                   & $<2$                   &  --                    & $<1$                  & --                  &  $<2$                \\
\ptw       &  --                   & $<\SI{350}{\GeV}$      &  --                    & --                    & $>\SI{250}{\GeV}$   &  $>\SI{250}{\GeV}$   \\
\mTW       &  --                   & --                     &  --                    & $<\SI{200}{\GeV}$     & --                  &  --                  \\
\ptlj      &  --                   & --                     &  $>\SI{70}{\GeV}$      & $>\SI{70}{\GeV}$      & $<\SI{350}{\GeV}$   &  $<\SI{350}{\GeV}$   \\
\bottomrule
\end{tabular}}
\egroup
\label{tab:CRs}
\end{table}


\printbibliography

\clearpage
\input{atlas_authlist}

\end{document}

%% file: atlas_authlist.tex
 
\begin{flushleft}
\hypersetup{urlcolor=black}
{\Large The ATLAS Collaboration}

\bigskip

\AtlasOrcid[0000-0002-6665-4934]{G.~Aad}$^\textrm{\scriptsize 102}$,
\AtlasOrcid[0000-0002-5888-2734]{B.~Abbott}$^\textrm{\scriptsize 120}$,
\AtlasOrcid[0000-0002-1002-1652]{K.~Abeling}$^\textrm{\scriptsize 55}$,
\AtlasOrcid[0000-0001-5763-2760]{N.J.~Abicht}$^\textrm{\scriptsize 49}$,
\AtlasOrcid[0000-0002-8496-9294]{S.H.~Abidi}$^\textrm{\scriptsize 29}$,
\AtlasOrcid[0000-0002-9987-2292]{A.~Aboulhorma}$^\textrm{\scriptsize 35e}$,
\AtlasOrcid[0000-0001-5329-6640]{H.~Abramowicz}$^\textrm{\scriptsize 151}$,
\AtlasOrcid[0000-0002-1599-2896]{H.~Abreu}$^\textrm{\scriptsize 150}$,
\AtlasOrcid[0000-0003-0403-3697]{Y.~Abulaiti}$^\textrm{\scriptsize 117}$,
\AtlasOrcid[0000-0002-8588-9157]{B.S.~Acharya}$^\textrm{\scriptsize 69a,69b,m}$
% ...
\bigskip
\\

% ...
$^{29}$Physics Department, Brookhaven National Laboratory, Upton NY; United States of America.\\
% ...
$^{35}$$^{(a)}$Facult\'e des Sciences Ain Chock, R\'eseau Universitaire de Physique des Hautes Energies - Universit\'e Hassan II, Casablanca;$^{(b)}$Facult\'{e} des Sciences, Universit\'{e} Ibn-Tofail, K\'{e}nitra;$^{(c)}$Facult\'e des Sciences Semlalia, Universit\'e Cadi Ayyad, LPHEA-Marrakech;$^{(d)}$LPMR, Facult\'e des Sciences, Universit\'e Mohamed Premier, Oujda;$^{(e)}$Facult\'e des sciences, Universit\'e Mohammed V, Rabat;$^{(f)}$Institute of Applied Physics, Mohammed VI Polytechnic University, Ben Guerir; Morocco.\\
% ...
$^{49}$Fakult\"{a}t Physik , Technische Universit{\"a}t Dortmund, Dortmund; Germany.\\
% ...
$^{55}$II. Physikalisches Institut, Georg-August-Universit\"{a}t G\"ottingen, G\"ottingen; Germany.\\
% ...
$^{69}$$^{(a)}$INFN Gruppo Collegato di Udine, Sezione di Trieste, Udine;$^{(b)}$ICTP, Trieste;$^{(c)}$Dipartimento Politecnico di Ingegneria e Architettura, Universit\`a di Udine, Udine; Italy.\\
% ...
$^{102}$CPPM, Aix-Marseille Universit\'e, CNRS/IN2P3, Marseille; France.\\
% ...
$^{117}$Department of Physics, New York University, New York NY; United States of America.\\
% ...
$^{120}$Homer L. Dodge Department of Physics and Astronomy, University of Oklahoma, Norman OK; United States of America.\\
% ...
$^{150}$Department of Physics, Technion, Israel Institute of Technology, Haifa; Israel.\\
$^{151}$Raymond and Beverly Sackler School of Physics and Astronomy, Tel Aviv University, Tel Aviv; Israel.\\
% ...
$^{a}$ Also Affiliated with an institute covered by a cooperation agreement with CERN.\\
$^{b}$ Also at An-Najah National University, Nablus; Palestine.\\
$^{c}$ Also at Borough of Manhattan Community College, City University of New York, New York NY; United States of America.\\
$^{d}$ Also at Center for High Energy Physics, Peking University; China.\\
$^{e}$ Also at Center for Interdisciplinary Research and Innovation (CIRI-AUTH), Thessaloniki; Greece.\\
$^{f}$ Also at Centro Studi e Ricerche Enrico Fermi; Italy.\\
% ...
$^{m}$ Also at Department of Physics, King's College London, London; United Kingdom.\\
% ...

\end{flushleft}
